\documentclass[11pt]{article}
\usepackage[T1]{fontenc}
\usepackage[utf8]{inputenc}
\usepackage{authblk}
\usepackage{bbding} 

\usepackage{graphicx,float,color}
\usepackage{amsmath}
\usepackage{amssymb}
\usepackage{amsxtra}
\usepackage{amsfonts}
\usepackage{hyperref}
\hypersetup{colorlinks = true, linkcolor = blue, urlcolor  = blue, citecolor = blue, anchorcolor = blue}
\usepackage{multirow}
\usepackage[numbers,sort&compress]{natbib}
\usepackage{epic}
\usepackage{eepic}
\usepackage{epsfig}
\usepackage{latexsym}
\usepackage{url}  
\usepackage{caption}
\usepackage[caption=false]{subfig}
\usepackage{bm}
\usepackage{geometry}
\usepackage{slashed}
\usepackage{simplewick}
\usepackage[vcentermath]{youngtab}
\usepackage{extarrows}
\makeatletter
\newcommand{\xRightarrow}[2][]{\ext@arrow 0359\Rightarrowfill@{#1}{#2}}
\makeatother

\title{Unified Pati–Salam from Noncommutative Geometry: \\
Overview and Phenomenological Remarks
}
\author{\vspace{0.1cm} Ufuk Aydemir \\ \small Department of Physics, Middle East Technical University, Ankara 06800, T\"urkiye, \\\small uaydemir@metu.edu.tr}

\begin{document}
\maketitle

\begin{abstract}
 The lack of clear new-physics signals at the LHC searches motivates models that can guide current and future collider searches. The spectral action principle within the noncommutative geometry (NCG) framework yields such models with distinctive phenomenology. This formalism derives the actions of the Standard Model, General Relativity, and beyond from the underlying algebra, putting them on a common geometric footing. Certain versions of Pati-Salam (PS) models with \textit{gauge coupling unification} and limited scalar content can be derived from an appropriate noncommutative algebra. In this paper\footnote{This contribution was prepared for the Proceedings of the 28th Bled Workshop, "What Comes Beyond the Standard Models", Bled, Slovenia, July 6 - 17, 2025.}, I review these gauge-coupling-unified Pati-Salam models and discuss their phenomenological aspects, focusing on the $S_1$ scalar leptoquark.
\\
\\
\textit{Keywords:} Noncommutative Geometry, Spectral Action, Pati-Salam, $R_{D^{(*)}}$ anomaly, scalar leptoquarks
\end{abstract}

\section{Introduction\label{sec:intro}}

%Although there have been several reported discrepancies between the theory and data, such as $B$-flavor anomalies etc (refs).  The long-existing discrepancy of the magnetic moment of the muon (g-2,  or $a_\mu$, anomaly) (ref) appears to be finally resolved (ref). There also exists a peculiar $W$ boson mass discrepancy, announced by Fermilab(ref), that contradicts all the other experiment including their first measurement. 

Since the discovery of the Higgs boson, a relentless effort has been put into the search for new physics beyond the Standard Model (SM). Contrary to high expectations stemming from the paradigms that contributed to the outstanding success of the SM, there has been no discovery of new physics at the LHC yet. In these difficult times, we should leave no stone unturned and explore all promising models that can help guide the experimental searches. In particular, models based on paradigms toward a deeper understanding of nature are specifically important. This is where Noncommutative Geometry (NCG)~\cite{Connes:1994yd,Connes:2007book} comes in. In analogy with quantum mechanics, NCG, by redefining notions of geometry, describes nature in terms of operator algebras instead of point sets of ordinary geometry. In the modern version of the framework, one can derive the SM, General Relativity (GR), and beyond by utilizing an appropriate action based on the Spectral Action principle~\cite{Chamseddine:1996rw,Chamseddine:1996zu,vanSuijlekom:2024jvw}. Reconciling the SM and gravity in a geometric setting, NCG offers a unifying picture of their origin.

The main object in NCG is the \textit{spectral triple} $(\mathcal{A}, \mathcal{H}, \mathcal{D})$, where $\mathcal{A}$ is an involutive algebra, $\mathcal{H}$ is a Hilbert space  on which the algebra acts as bounded operators, and the (generalized) Dirac operator $\mathcal{D}$, a (possibly) unbounded self-adjoint operator. The spectral triple is augmented by extra structures~\cite{Chamseddine:1996rw} such as a $\mathbb{Z}_2$ grading through the chirality operator $\Gamma$ (generalized $\gamma_5$) and an antilinear unitary operator $\mathcal{J}$ (generalized charged conjugation) on $\mathcal{H}$. This structure encodes the information on the geometry; the ordinary points are now replaced by the spectrum of the Dirac operator, the inverse of which acts as a metric. On the other hand, the information on the manifold is recovered by the algebra $\mathcal{A}$. 

The spectral data $(\mathcal{A}, \mathcal{H}, \mathcal{D},\Gamma,\mathcal{J})$ is given as the product of the ordinary part, corresponding to four-dimensional manifold $M$, with a finite space with noncommutative geometry. This corresponds to two sheeted spacetime with $M\times \mathbb{Z}_2$. One can obtain physical models, depending on the choice of the finite part, and using the spectral action given by $\mathcal{S}=\mathcal{S}_F+\mathcal{S}_B=\left(J \psi, \mathcal{D}_A \psi\right)+\operatorname{Tr}\left[\chi(\mathcal{D}_A/\Lambda)\right]$, where the former term corresponds to the fermionic sector (which also yields the Yukawa terms due to the Higgs embedded in the Dirac operator) and the latter is for purely bosonic part~\cite{Chamseddine:1996rw,Chamseddine:1996zu}. "$\operatorname{Tr}$" is the trace over the Hilbert space $\mathcal{H}$ . The cutoff function $\chi$ acts as a regulator that selects the eigenvalues of the covariant Dirac operator, $\mathcal{D}_A$, smaller than the cutoff $\Lambda$. 

In the basic construction~\cite{Chamseddine:1996rw,Chamseddine:1996zu}, the algebra is chosen as $\mathcal{A}=C^{\infty}(M) \otimes \mathcal{A}_F$ such that the finite part of the algebra is given as $\mathcal{A}_F=\mathbb{C} \oplus \mathbb{H} \oplus M_3(\mathbb{C})$, where $\mathbb{H} \subset M_2(\mathbb{C})$ is the algebra of quaternions, and $M_3(\mathbb{C})$ is the algebra of $3 \times 3$ matrices with elements in $\mathbb{C}$. Then, the spectral action yields the SM action and the action of a modified gravity model, the latter of which consists of the Einstein-Hilbert and the cosmological constant terms, a non-minimal coupling term between the Higgs boson and the curvature, the Gauss-Bonnet term, and the Weyl (or the conformal gravity) term. The SM parameters are included in the Dirac operator, and the Higgs boson arises as the connection in the extra discrete dimension. The gauge transformations emerge from the unitary inner automorphisms of the algebra $\mathcal{A}$ while diffeomorphisms arise from the outer automorphisms. In Refs.~\cite{Chamseddine:2013rta,Chamseddine:2015ata}, by utilizing the algebra $\mathcal{A}_F=\mathbb{H}_R \oplus \mathbb{H}_L \oplus M_4(\mathbb{C})$, constructed models with Pati-Salam (PS) gauge structure $G_{422}=SU(4)\times SU(2)_L \times SU(2)_R$. Depending on whether the so-called order-one condition is satisfied, three versions of these models are obtained with different scalar content and with/without left-right symmetry. 

PS models based on noncommutative geometry (NCG-PS) come with a number of appealing features compared to the ordinary counterparts, heavily studied in the literature~\cite{Pati:1974yy,Mohapatra:1974gc,Mohapatra:1974hk}. First, NCG-PS  models require \textit{gauge coupling unification}, which is a feature that is not mandatory in ordinary PS models unless they are embedded in a larger group such as $SO(10)$~\cite{Chang:1984qr,Bertolini:2009qj,Aydemir:2016qqj,Aydemir:2019ynb,Aydemir:2022lrq}. Furthermore, as opposed to the ordinary PS models, NCG-PS models come with a restricted content of scalar fields with enough number and quality required for certain symmetry-breaking patterns and mass generation; not all interaction terms are allowed in the Lagrangian. These features increase the predictivity of these models. Finally, the proton stability due to the light leptoquarks $S_1$ (which is our interest in our work, as will be discussed below) is not generally guaranteed; in NCG-PS, on the other hand, the diquark couplings of some of these leptoquarks are missing, and them being light does not cause an issue of concern, as pointed out in Ref.~\cite{Aydemir:2018cbb}. I also note that the fermion content in the NCG-PS models is the same as the Standard Model (SM) plus the right-handed neutrinos of each generation, similar to the GUT models.

In this paper, I briefly review NCG-based Pati-Salam (NCG-PS) models, with gauge coupling unification, and discuss certain differences from the ordinary Pati-Salam models. Regarding the low-energy phenomenology, I focus on the new physics scenario of  TeV-scale leptoquark(s) of $S_1$ type.

%%%%%%%%%%%%%%%%%%%%%%%%%%%%%%%%%%%%%%%%%%%%%%%%%%%%%%%%%%%%
\section{Minimal NCG framework;\\ Spectral Standard Model (with gravity)}
\label{sec:NCG}
%%%%%%%%%%%%%%%%%%%%%%%%%%%%%%%%%%%%%%%%%%%%%%%%%%%%%%%%%
As mentioned above, the algebra chosen for the construction that accommodates the SM is given as~\cite{Chamseddine:1996rw,Chamseddine:1996zu,vanSuijlekom:2024jvw} 
\begin{equation}
\label{AlgebraSM}
\mathcal{A}=C^{\infty}(M) \otimes \mathcal{A}_F\;,\quad \mathrm{where} \quad \mathcal{A}_F=\mathbb{C} \oplus \mathbb{H} \oplus M_3(\mathbb{C})\;,
\end{equation}
$\mathbb{H} \subset M_2(\mathbb{C})$ is the algebra of quaternions, and $M_3(\mathbb{C})$ is the algebra of $3 \times 3$ matrices with elements in $\mathbb{C}$. One can easily recognize the correspondence between the elements of the finite algebra $\mathcal{A}_F$ and the SM gauge groups $U(1)$, $SU(2)$, and $S(3)$. The action, called the \textit{spectral action}, is constructed as 
\begin{equation}
\label{actiontotal}
\mathcal{S}=\mathcal{S}_F+\mathcal{S}_B=\left(J \psi, \mathcal{D}_A \psi\right)+\operatorname{Tr}\left[\chi\left(\frac{\mathcal{D}_A}{\Lambda}\right)\right]\;.
\end{equation}
The second term corresponds to the purely bosonic part of the action, whereas the first term is the fermionic part, including the Yukawa sector. Note that the Dirac operator includes the Higgs field as the gauge connection between the two sides of the finite part of the spacetime (two points, at each of which a 4d manifold is located) connecting the left and right sectors. $\chi$ is the cutoff function that selects the eigenvalues of the covariant Dirac operator $\mathcal{D}_A$ smaller than the cutoff scale $\Lambda$. $J$ is the generalized charge conjugation to account for the antiparticles and manages the real structure on $\mathcal{H}$~\cite{Chamseddine:1996rw,Chamseddine:1996zu}; it is an antiunitary operator acting on the Hilbert space $\mathcal{H}$. Finally; the covariant Dirac operator, $\mathcal{D}_A$, in terms of the unperturbed Dirac operator $\mathcal {D}$, real structure operator $J$,  and a Hermitian one-form potential $A$, is given as
\begin{equation}
\mathcal{D}_A=\mathcal{D}+A+J A J^{\dagger}, \quad \mathrm{where}\quad A=\sum a_i\left[\mathcal{D}, b_i\right], \quad a_i, b_i \in \mathcal{A}, \quad\mathrm{and}\quad A=A^*\;.
\end{equation}
The first equation accounts for the inner fluctuations of the line element (inverse $D$). There is also the so-called first-order condition~\cite{Chamseddine:2013kza}, given as 
\begin{equation}
\left[[\mathcal{D}, a], J b J^{-1}\right]=0, \quad \forall a, b \in \mathcal{A}\;,
\end{equation}
which is specifically important beyond the standard framework, as will be mentioned in the next section.

The generalized Dirac operator is chosen accordingly to yield the fermionic sector of the SM, including the Yukawa terms. The construction predicts certain relations among the Yukawa couplings to be satisfied at the energy scale at which the spectral action is assumed to emerge~\cite{Chamseddine:1996rw,Chamseddine:1996zu}. Note also that this construction requires gauge coupling unification with the same field content as the SM. At this point, we need to look at the bosonic part of the action, given in Eq.~(\ref{actiontotal}). Since the details are not important for this talk (and the interested reader can check Refs.~\cite{Chamseddine:1996rw,Chamseddine:1996zu,vanSuijlekom:2024jvw} for more details), I just give the final action here. The bosonic part of the spectral SM action is given as 
\begin{equation}
\label{bosonicaction}
\begin{aligned}
\mathcal{S}_B & =\int\left(\frac{1}{2 \kappa_0^2} R+\alpha_0 C_{\mu \nu \rho \sigma} C^{\mu \nu \rho \sigma}+\gamma_0+\tau_0 R^* R^*\right. \\
& +\frac{f_0}{2 \pi^2}\left[g_3^2 G_{\mu \nu}^i G^{\mu \nu i}+g_2^2 F_{\mu \nu}^m F^{\mu \nu m}+\frac{5}{3} g_1^2 B_{\mu \nu} B^{\mu \nu}\right] \\
& \left.+\left|D_\mu H\right|^2-\mu_0^2|H|^2-\xi_0 R|H|^2+\lambda_0|H|^4+O\left(\frac{1}{\Lambda^2}\right)\right) \sqrt{|g|}\; d^4 x\;,
\end{aligned}
\end{equation}
where $R$ is the usual Ricci scalar, $C_{\mu \nu \rho \sigma}$ is the Weyl tensor, $R^* R^*$ denotes the topological Gauss-Bonnet term, and $\left(\kappa_0, \alpha_0, \gamma_0, \tau_0, \mu_0, \xi_0, \lambda_0\right)$ are constants defined in terms of original parameters in the theory. The theory is truncated in the lowest relevant order in the energy scale $\Lambda$ and is clearly not UV-complete in the common sense, which can be viewed as a shortcoming of the framework. Another issue is that a QFT formalism that is faithful to the NCG structure has not been completely established (see Ref.~\cite{vanNuland:2021irt} for a recent study to this end.) Therefore, it is conceivable to consider that the action emerges as a geometric structure at a certain high energy scale, much lower than the cutoff scale $\Lambda$. On top of that geometric structure, we assume that the usual QFT framework is valid as an initial approximation. Note that the model based on Lagrangian (\ref{bosonicaction}) leads to the wrong Higgs mass. In Ref.~\cite{Chamseddine:2012sw}, the authors argue that there is an extra singlet field in the theory that can correct the Higgs mass.

One can see in the bosonic action, given in Eq.~(\ref{bosonicaction}), the requirement of gauge coupling unification. Sticking to the canonical normalization of the kinetic terms, one obtains the condition
\begin{equation}
g_3^2=g_2^2=\frac{5}{3} g_1^2\;,
\end{equation}
assumed to satisfy at a certain unification scale $M_U$. One may have an immediate tendency to identify $M_U$ with the cutoff scale $\Lambda$ ($M_U\sim \Lambda$), but for the truncation of the action in (\ref{bosonicaction}) to be acceptable, it is more reasonable to require $M_U\ll \Lambda$. Since in this minimal construction, we only have the SM field content, the gauge coupling unification is not achievable with the usual perturbative RG running. Ref.~\cite{Kurkov:2017wmx} examined a modification of the minimal construction that yields several additional scalar fields; however, Ref.~\cite{Aydemir:2019txw} demonstrated that gauge coupling unification cannot be realized within this modified framework, regardless of the mass hierarchy of the extra fields. Therefore, it is necessary to extend the spectral formalism beyond the minimal framework. 

\section{Pati-Salam models from NCG with gauge coupling unification}
\label{sec:models}
%%%%%%%%%%%%%%%%%%%%%%%%%%%%%%%%%%%%%%%%%%%%%%%%%%%%%%%%%
\subsection{Basics}

In Refs.~\cite{Chamseddine:2013rta,Chamseddine:2015ata}, the formalism was extended by changing the the algebra, given in Eq.~(\ref{AlgebraSM}), to 
\begin{equation}
\label{AlgebraPS}
\mathcal{A}=C^{\infty}(M) \otimes \mathcal{A}_F\;,\quad \mathrm{where} \quad \mathcal{A}_F=\mathbb{H}_R \oplus \mathbb{H}_L \oplus M_4(\mathbb{C})\;,
\end{equation}
 and selecting the rest of the spectral data appropriately.  The framework yields three models with different scalar contents and initial gauge symmetries, depending on whether the order-one condition is fully satisfied. The scalar contents of the models are listed in Table~\ref{NCG-HiggsContent}, with the notation $G_{422D}  =  SU(4)_C\otimes SU(2)_L\otimes SU(2)_R\otimes D$, where the $D$ symbol refers to the left-right symmetry, a $Z_2$ symmetry which keeps the left and the right sectors equivalent. The symbol $G_{422}$ is used for the case where the Pati-Salam gauge group appears without the $D$ symmetry. For the full spectral Pati-Salam action, including gravitational terms, see Ref.~\cite{Chamseddine:2013rta}. 
%Note also that we adopt the normalization of the hypercharge $Y$ so that $Q_{em} = I_3^L + Y = I_3^L + I_3^R + (B-L)/2$.

%%%%%%%%%%%%%%%%%%%%%%%%%%%%%%%%%%%%%%%%%%%%%%%%%%%%%%%%%%%%%
\begin{table}[ht!]
\caption{The scalar content of the three NCG-based Pati-Salam models.
}
{\begin{tabular}{l|l|ll}
\hline
{\small Model} & {\small Symmetry} & {\small Scalar Content} \vphantom{\Big|}\\
\hline\hline
A & $G_{422}$  & $\phi(1,2,2)_{422}$, $\Sigma(15,1,1)_{422}$, & \hspace{-3mm}$\widetilde{\Delta}_R(4,1,2)_{422}$  \vphantom{\bigg|}\\
\hline
B & $G_{422}$  & $\phi(1,2,2)_{422}$, $\widetilde{\Sigma}(15,2,2)_{422}$, & \hspace{-3mm}$\Delta_R(10,1,3)_{422}$, $H_R(6,1,1)_{422}$ \vphantom{\bigg|}\\
\hline
 C & $G_{422D}$ & $\phi(1,2,2)_{422}$, $\widetilde{\Sigma}(15,2,2)_{422}$, & \hspace{-3mm}$\Delta_R(10,1,3)_{422}$, $H_R(6,1,1)_{422}$, \vphantom{\bigg|}\\
  & & & \hspace{-3mm}$\Delta_L(10,3,1)_{422}$, $H_L(6,1,1)_{422}$ \vphantom{\Big|}\\
\hline\end{tabular}}
\label{NCG-HiggsContent}
\end{table}
%%%%%%%%%%%%%%%%%%%%%%%%%%%%%%%%%%%%%%%%%%%%%%%%%%%%%%%%%%%%%

A common way to break the PS symmetry into the SM can be schematically shown as~\cite{Aydemir:2018cbb}
\begin{equation}
\label{chain}
\begin{array}{lll}
\mathrm{NCG} &\;\xRightarrow{\;\;\;M_U\;\;\;}\;\; G_{422D} &\underset{\langle\Delta_R\rangle}{\xrightarrow{\quad M_C\quad}}\;\; G_{321}\;,
\end{array}
\end{equation}
where the double arrow denotes the symmetry emerging from the underlying NCG at the unification scale $M_U$ (or above), while the single arrow denotes the spontaneous symmetry breaking in the usual way. Breaking of the Pati-Salam symmetry into the SM proceeds through the SM singlet within $\Delta_R(10,1,3)_{422}$, acquiring a VEV. Depending on whether the selected model (i.e. A, B, or C) contains the necessary fields,  intermediate symmetry-breaking stages can be included~\cite{Aydemir:2015nfa, Aydemir:2016xtj}.

As in the case of ordinary PS models, the fermions are in $(4,2,1)_{422}$ and $(4,1,2)_{422}$ representations, which can be put in the following form.
\begin{eqnarray}
\psi_{aI} & = & (\psi_{a0},\psi_{ai}) \;=\; 
\left( \begin{array}{ll} \psi_{10}, & \psi_{1i} \\ \psi_{20}, & \psi_{2i} \end{array} \right) \;=\;
%\psi_L \;=\;
\left( L_L, Q_L \right) \;=\;
\left( \begin{array}{ll} \nu_L, & u_L \\ e_L, & d_L \end{array} \right)
\;,\cr
\psi_{\dot{a}I} & = & (\psi_{\dot{a}0},\psi_{\dot{a}i}) \;=\; 
\left( \begin{array}{ll} \psi_{\dot{1}0}, & \psi_{\dot{1}i} \\ \psi_{\dot{2}0}, & \psi_{\dot{2}i} \end{array} \right) \;=\;
%\psi_R \;=\;
\left( L_R, Q_R \right) \;=\;
\left( \begin{array}{ll} \nu_R, & u_R \\ e_R, & d_R \end{array} \right)
\;,
\end{eqnarray}
which is the SM fermion content with the right-handed neutrinos for each generation. The dotted and undotted lower-case Latin letters toward the beginning of the alphabet denote $SU(2)_R$ and $SU(2)_L$ indices in the fundamental representation, respectively: e.g. $\dot{a} = 1,2$ and $a = 1,2$. The $SU(4)$ index in the fundamental representation
is denoted with upper-case Latin letters toward the middle of the alphabet: e.g. $I=0,1,2,3$,
where $I=0$ is the lepton index and $I=i=1,2,3$ are the quark-color indices. The spinor and generation indices are omitted. Complex (hermitian, Dirac) conjugation raises or lowers both indices, e.g.
\begin{equation}
\overline{\psi}{}^{aI} \;=\; \overline{\psi_{aI}} \;,\qquad
\overline{\psi}{}^{\dot{a}I} \;=\; \overline{\psi_{\dot{a}I}} \;.
\end{equation}
In the case of the $SU(2)$'s, the index can be lowered or raised using
\begin{equation}
(\epsilon)_{ab}\;,\qquad
(\epsilon^\dagger)^{ab}\;,\qquad
(\epsilon)_{\dot{a}\dot{b}}\;,\qquad
(\epsilon^\dagger)^{\dot{a}\dot{b}}\;,
\end{equation}
where $\epsilon = i\sigma_2$. 

The complex scalar fields in this framework are given as  
%$\Sigma^{bJ}_{\dot{a}I}$, $H_{\dot{a}I \dot{b}J}$, and $H_{aIbJ}^{\phantom{\dagger}}$ can, 
%in general, consist of the following $G_{422}$ representations:
%
\begin{eqnarray}
\label{scalarfields}
\Sigma^{bJ}_{\dot{a}I}
& = & (1,2,2)_{422} + (15,2,2)_{422} \;, \vphantom{\Big|}\cr
H_{aIbJ}^{\phantom{\dagger}} & = & (6,1,1)_{422} + (10,3,1)_{422} \;, \vphantom{\Big|}\cr
H_{\dot{a}I \dot{b}J} & = & (6,1,1)_{422} + (10,1,3)_{422} \vphantom{\Big|}\;. 
\end{eqnarray}
In model C, we have all of these fields, whereas in model B, which is, unlike model C, is not let-right symmetric, $H_{aIbJ}^{\phantom{\dagger}} $ is turned off. Finally, in model A, which is referred to as the \textit{composite} model in Refs.~\cite{Chamseddine:2013rta,Chamseddine:2015ata}, the $H_{\dot{a}I \dot{b}J}$ and $\Sigma^{bJ}_{\dot{a}I}$ fields are not fundamental and composed of the fields $\phi(1,2,2)_{422}$, $\Sigma(15,1,1)_{422}$, and $\widetilde{\Delta}_R(4,1,2)_{422}$.  
%Again, we are interested in only model C, so we have the scalars $H_{aIbJ}^{\phantom{\dagger}}$, $H_{\dot{a}I \dot{b}J} $, and $\Sigma^{bJ}_{\dot{a}I}$ with the decomposition given in Eq.~(\ref{scalarfields}).    

\subsection{Remarks on the low energy phenomenology}

For the sake of this presentation, I will continue with the most general model, model C.  I will only focus on the Yukawa sector, leaving out the scalar sector since the latter is not relevant for our discussion. The  $G_{422D}$ invariant Yukawa terms for each family of fermions can be written schematically as~\cite{Chamseddine:2013rta,Aydemir:2018cbb}
\begin{equation}
\label{eq:ncg-yukawa}
\mathcal{L}_\mathrm{Y}
\;=\;
\Bigl(\,
\overline{\psi}{}^{\dot{a}I}
\gamma_5 \Sigma^{bJ}_{\dot{a}I\vphantom{\dot{b}}} 
\psi_{bJ\vphantom{\dot{b}}}^{\vphantom{C}}
+ \overline{\psi^C}{}_{aI\vphantom{\dot{b}}}\gamma_5 H^{aIbJ} \psi_{bJ\vphantom{\dot{b}}}^{\vphantom{C}}
+ \overline{\psi^C}{}_{\dot{a}I\vphantom{\dot{b}}}\gamma_5 H^{\dot{a}I \dot{b}J} \psi_{\dot{b}J}^{\vphantom{C}}
\,\Bigr)
\;+\; h.c.\;,
\end{equation}
where $\psi^C = C\overline{\psi}^T$. The Yukawa coupling constants are embedded in the complex scalar fields $\Sigma_{\dot{a} I}^{b J}, H^{a I b J}$, and $H^{\dot{a} I \dot{b} J}$. The $\gamma_5$ that appears in this expression is due to the grading of the geometry. Actually, the "grading" in the product space $M\times F$, where $M$ is the continuous 4D manifold and $F$ is the finite part, is realized by the generalized chirality operator $\Gamma=\gamma_5 \otimes \gamma_F$; here, $\gamma_5$ is the usual chirality operator for the continuous manifold and $\gamma_F$ introduces a $Z_2$ grading and responsible for the algebra of quaternionic matrices $M_2 (\mathbb{H})$  into $\mathbb{H}_R \oplus \mathbb{H}_L$~\cite{Chamseddine:2019fjq}. The first term in Eq.~(\ref{eq:ncg-yukawa}) yields terms containing fermions with opposite chiralities (LR) through the `connection' $\Sigma_{\dot{a} I}^{b J}$. The LL and RR terms arise from the second and third terms due to fields $H^{a I b J}$ and $H^{\dot{a} I \dot{b} J}$, respectively, which connect fermions and antifermions with the same chirality.  

%In Eq.~(\ref{eq:ncg-yukawa}), the `connection' $\Sigma_{\dot{a} I}^{b J}$  

%$H^{a I b J}$

After spontaneous symmetry breaking to the SM, the terms in Eq.~(\ref{eq:ncg-yukawa}) yield all the SM terms in addition to interaction terms between the SM sector and new fields. This is similar to the ordinary Pati-Salam models. However, in addition to gauge coupling unification and the restricted scalar content, the NCG-based Pati–Salam (NCG–PS) models differ further from the ordinary ones in terms of the allowed Lagrangian terms. Some of the terms that would be expected in the ordinary case (unless some extra ad-hoc symmetries are imposed) do not appear in the NCG-PS models automatically due to the underlying noncommutative geometry. These differences lead to advantages/predictions that could help to probe these models. 

\begin{figure}[t]
\begin{center}
\includegraphics[width=12cm]{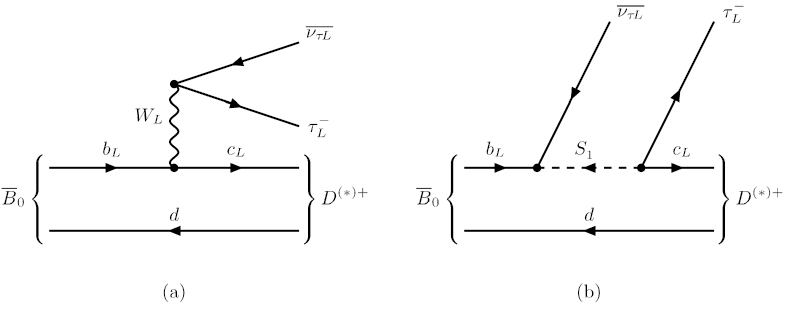}
\caption{\label{BtoDlnu}
(a) SM and (b) $S_1$ leptoquark contribution to 
$\overline{B}_0\to D^{(*)+}\tau^-\overline{\nu_{\tau\!L}}$. Adapted from Ref.~\cite{Aydemir:2018cbb}.}
\end{center}
\end{figure}

Consider the anomalies in the charged-current $B$ decays that have been around for over a decade~\cite{BaBar:2012obs,LHCb:2015gmp,Belle:2017ilt}\footnote{See Ref.~\cite{Crivellin:2025txc} for a recent review of current anomalies in high energy physics.}, notably in the  $R_{D^{(*)}}$ observables, defined as
\begin{equation}
R_{D^{(*)}}=\frac{\mathrm{BR}\left(B \rightarrow D^{(*)} \tau \nu\right)}{\mathrm{BR}\left(B \rightarrow D^{(*)} \ell \nu\right)}\;,
\end{equation}
where $\ell=\{e, \mu\}$ and $\mathrm{BR}$ denotes branching ratio. The latest averaged experimental values of these observables deviate from the SM values by more than $3 \sigma$, as estimated by the Heavy Flavor Averaging Group~\cite{HeavyFlavorAveragingGroupHFLAV:2024ctg}. The leptoquark $S_1(\overline{3},1,1 / 3)_{321}$ is a popular minimal option, providing a tree-level solution~\cite{Bauer:2015knc,Angelescu:2018tyl,Aydemir:2019ynb,Aydemir:2022lrq,Aydemir:2018cbb,Aydemir:2023sty} (see Fig.~\ref{BtoDlnu}).\footnote{The leptoquark $S_3\left(\overline{3}, 3, \frac{1}{3}\right)_{321}$ can also provide a solution~\cite{Angelescu:2018tyl} through its component with electric charge $1/3$. While it provides a safe option regarding the proton stability, it comes in rather larger multiplets compared to $S_1(\overline{3},1,1 / 3)_{321}$, which might be more appealing for a compact explanation.}

%%%%%%%%%%%%%%%%%%%%%%%%%%%%%%%%%%%%%%%%%%%%%%%%%%%%%%%%%%%%%%%%%%%%%%%%%%%%%%%%
%%%%%%%%%%%%%%%%%%%%%%%%%%%%%%%%%%%%%%%%%%%%%%%%%%%%%%%%%%%%%%%%%%%%%%%%%%%%%%%%

The existence of such a scalar leptoquark at the TeV scale would prompt important questions. The immediate one is its UV origin. Leptoquarks~\cite{Buchmuller:1986zs,Dorsner:2016wpm} appear in supersymmetric extensions of the SM, composite (strongly-coupled) models, grand unified theories, and Pati-Salam-type partially unified models, and finally our NCG-PS models with complete gauge coupling unification.  NCG-PS models have an appealing feature over the other.  As previously mentioned, due to the underlying noncommutative structure, the scalar sector of each of the three models is quite restrictive, thus predictive. Another important point would be the mechanism that prevents diquark couplings of $S_1$ type leptoquark, since such couplings would mediate proton decay~\cite{Nath:2006ut}. In fact, proton stability is the reason why, in grand unified theories, $S_1$ type leptoquarks are assumed to be heavy near the unification scale. Particularly in $SO(5)$ and $SO(10)$ theories, as well as in supersymmetric theories in general, the SM Higgs doublet is accompanied by a leptoquark in a larger multiplet. Keeping the leptoquark heavy while having the SM doublet light is known as the infamous double-triplet splitting problem. 

\begin{figure}[ht]
\begin{center}
\includegraphics[width=10cm]{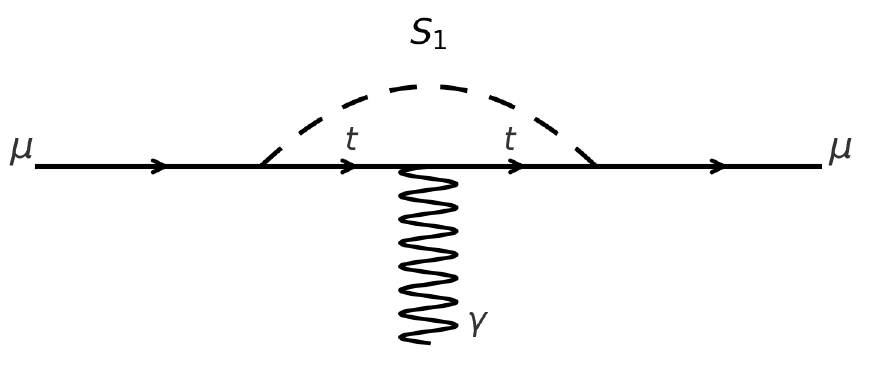}
\caption{\label{amu}
Leading order contribution to $a_{\mu}$ from a $S_1$ leptoquark.}
\end{center}
\end{figure}

 In NCG-PS models, even though there are several $S_1$ type leptoquarks, only one of them can appropriately provide a solution to the  $R_{D^{(*)}}$ anomaly,  reflecting the restricted and predictive aspect of this framework. The $S_1$ leptoquarks in models A and B couple either only to right-handed fermions or only to diquarks~\cite{Aydemir:2018cbb}; hence, they are not useful for the  $R_{D^{(*)}}$  anomaly.  On the other hand, in model C,  one of the leptoquarks in a (complexified) $ H(6,1,1)_{422}$ possesses the required couplings to left-handed fermions, while lacking the diquark couplings. Therefore, it can provide a solution and does not mediate proton decay.  
 Namely, in model C, we have the object
 \begin{equation}
H_{a I b J}=\Delta_{(a b)(I J)}+H_{[a b][I J]}=\Delta_L(10,3,1)_{422}+H_L(6,1,1)_{422}\;.
 \end{equation}
 The sextet is decomposed into the SM components as
 \begin{equation}
 \label{decomposition1}
H_L(6,1,1)_{422}=H_{3 L}\left(3,1,-\frac{1}{3}\right)_{321}+H_{\overline{3}  L}\left(\overline{3}, 1, \frac{1}{3}\right)_{321}\;,
 \end{equation}
 which, for the complexified sextet, corresponds to two different leptoquarks. Then, the second term (with its $h.c.$) in Eq.~(\ref{eq:ncg-yukawa}) yields, for each family, the terms~\cite{Aydemir:2018cbb}
\begin{eqnarray}
\lefteqn{
\bar{\psi}_{a I}^C \gamma_5 H^{[a b][I J]} \psi_{b J}+\text { h.c. }
\vphantom{\Big|}
}\cr
& = & 2\biggl[
\Bigl(\overline{d_{Lj}^C}\nu_L^{\vphantom{\dagger}} - \overline{u_{Lj}^C}e_L^{\vphantom{\dagger}}\Bigr)
H_{3L}^{*j} 
+\;
\varepsilon^{ijk}\overline{u_{Li}^C}d_{Lj}^{\phantom{\dagger}}H_{\overline{3}Lk}^{*}
\biggr]
\;+\; h.c.\;,
\label{HLcouplings}
\end{eqnarray}
where $H_{3L}^{*}$ can be identified as the "good" leptoquark that has left-handed couplings while lacking diquark couplings. On the other hand,  $H_{\overline{3}L}^{*}$ couples to diquarks, thus mediating proton decay, and should be taken heavy.\footnote{ In the ordinary, non-unified, PS framework, the scalar sector is generally constructed with $\Delta_R(10,1,3)_{422} \text {, } \Delta_L(10,3,1)_{422}$, and the bidoublet $\phi(1,2,2)_{422}$. Due to the totally symmetric nature of the scalar sector, there exists a symmetry that prevents these couplings~\cite{Mohapatra:1980qe}. However, if one includes the multiplet $(6,1,1)_{422}$, as is the case in NCG-PS models, which transforms antisymmetrically under $SU(4)$, then the symmetry no longer naturally exists. In our case,  the proton-decay–mediating diquark couplings of our leptoquark are automatically absent due to the geometric construction. Compared to the regular PS models, NCG-PS may not seem to have an advantage regarding this issue, since in both cases, the proton is safe. Yet, considering other advantages of the latter, such as gauge coupling unification, restricted scalar content, and an underlying geometric explanation for the theory that reconciles also gravity on the same footing, one can argue in favor of the appeal of NCG-PS models.} The exclusive left-handed couplings of $S_1$ should not impede $R_{D^{(*)}}$ solutions; however, they reduce the available parameter space compared with regular models. One possible significance of the absence of the right-handed couplings could be in the context of the magnetic moment of the muon, $g-2$ ($a_\mu$). The $S_1$ contribution to $a_\mu$ (see Fig.~\ref{amu}) with only left-handed couplings is suppressed and comes with a negative sign. This would have been a problem for the $a_\mu$ discrepancy between theory and experiment that persisted for several decades. However, recent developments on both sides suggest that such a discrepancy may be absent.

%%%%%%%%%%%%%%%%%%%%%%%%%%%%%%%%%%%%%%%%%%%%%%%%%%%%%%%%%%%%%%%%%%%%%%%%%%%%%%%%
%%%%%%%%%%%%%%%%%%%%%%%%%%%%%%%%%%%%%%%%%%%%%%%%%%%%%%%%%%%%%%%%%%%%%%%%%%%%%%%%

Note that I have selected the above model just as a sample. I have aimed to address the  $R_{D^{(*)}}$  anomaly with a single $S_1$ that does not cause proton decay. The only one at our disposal was $H_{3L}^{*}$, contained in $H_L(6,1,1)_{422}$ in model C.  If the right-handed couplings of $S_1$ are also needed to address a measurement of a particular observable, one can include the one in $H_R(6,1,1)_{422}$ (see Table~\ref{NCG-HiggsContent}), whose decomposition is the counterpart of Eq.~(\ref{decomposition1}) with the components denoted as $H_{3 R}$ and $H_{\overline{3}R}$. The corresponding Yukawa terms, in addition to the ones in Eq.~(\ref{HLcouplings}), come from the third term (with its $h.c$) in Eq.~(\ref{eq:ncg-yukawa})  given for each family as~\cite{Aydemir:2018cbb}  
\begin{eqnarray}
\lefteqn{\overline{\psi^C}_{\dot{a}I\vphantom{\dot{b}}}\gamma_5 H^{[\dot{a}\dot{b}][IJ]} \psi_{\dot{b}J}
\;+\; h.c. 
\vphantom{\bigg|}
}\cr
& = & 2\biggl[
 \left(
 \overline{d_{Rj}^C}\nu_R^{\vphantom{\dagger}}
-\overline{u_{Rj}^C}e_R^{\vphantom{\dagger}}  
 \right) H_{3R}^{*j}
+\,
 \varepsilon^{ijk}
 \overline{u_{Ri}^C}d_{Rj}^{\vphantom{\dagger}} 
 H_{\overline{3}Rk}^{*} 
\biggr] \;+\; h.c.\;,
\label{HRcouplings}
\end{eqnarray}
with $H_{3R}^{*j}$ being the "good" leptoquark with no diquark couplings.

Besides $a_\mu$ and $R_{D^{(*)}}$, one can analyze

$$
R_{K^{(*)}} \equiv \frac{\mathrm{BR}\left(B \rightarrow K^{(*)} \mu^{+} \mu^{-}\right)}{\mathrm{BR}\left(B \rightarrow K^{(*)} e^{+} e^{-}\right)}\;\quad\mathrm{and}\quad R_K^\nu \equiv \frac{\mathrm{BR}(B \rightarrow K \nu \nu)}{\mathrm{BR}(B \rightarrow K \nu \nu)_{\mathrm{SM}}}\;,
$$
along with observables, such as $\operatorname{BR}(\tau \rightarrow \mu \gamma), \operatorname{BR}(\tau \rightarrow 3 \mu)$, $\operatorname{BR}\left(B_c \rightarrow \tau \nu\right)$, among others~\cite{Bhaskar:2022vgk,Aydemir:2022lrq}, for specific candidate particle(s) (e.g.~$S_1$). Given the restricted and predictive structure of NCG–PS models, such a global analysis can yield stringent, model-discriminating information on physics beyond the SM. 

%We have discussed a particular $S_1$ leptoquark contribution to Rd and amu to demonstrate the teh comparison of the NCG-PS models to the regular ones. One can investigate further topics such as two-Higgs-doublet models, baryogenesis, or models including other leptoquarks. Any physica phenomean or observed field would have direct implication regarding these models due to the their restricted scalar content and coupling structure, and hence making it predictive and therefore appealing. 

%%%%%%%%%%%%%%%%%%%%%%%%%%%%%%%%%%%%%%%%%%%%%%%%%%%%%%%%%%%%%%%%%%%%%%%
\section{Summary and Outlook}
\label{sec:Conc}

By analogy with quantum mechanics, noncommutative geometry (NCG) reformulates geometry in terms of operator algebras rather than point sets. In its modern formulation, an action based on the spectral action principle~\cite{Chamseddine:1996rw,Chamseddine:1996zu,vanSuijlekom:2024jvw} can be used to accommodate the SM, GR, and beyond. By placing the particle physics theories and gravity on the same geometric footing, this framework offers a unified perspective on their origin and may be viewed as a step toward quantum gravity.

Using an appropriate algebra, one can construct models with the Pati–Salam (PS) gauge group $SU(4)\times SU(2)_L \times SU(2)_R$ with the condition of gauge-coupling unification~\cite{Chamseddine:2013rta,Chamseddine:2015ata}. Depending on an underlying geometric condition, referred to as the order-one condition, three versions of these models are obtained with different scalar content and with/without left-right symmetry. 

The NCG–based Pati–Salam framework, with its restricted scalar content and Yukawa structure, yields a characteristic setup. As an illustration, I focused on a TeV-scale $S_1$ leptoquark. In Model C, the required couplings to address current flavor anomalies are present, whereas proton-decay–mediating diquark couplings of this leptoquark are automatically absent due to the geometric construction, rather than by ad hoc assumptions.

The absence of clear new-physics signals at the LHC motivates models that can guide current and future searches. The spectral action formalism in the NCG framework provides distinctive Pati-Salam models with gauge coupling unification that deserve attention.

\section*{Acknowledgements}
This work was supported by The Scientific and Technological Research Council of T\"urkiye (T\"UB\.ITAK) B\.{I}DEB 2232-A program under project No.~121C067.

%% The bibliography section


\begin{thebibliography}{99}
\bibitem{Connes:1994yd}
A.~Connes, \emph{{Noncommutative geometry}}.
\newblock 1994.

\bibitem{Connes:2007book}
A.~Connes and M.~Marcolli, \emph{{Noncommutative Geometry, Quantum Fields and
  Motives}}.
\newblock American Mathematical Society, 2007.

\bibitem{Chamseddine:1996rw}
A.~H. Chamseddine and A.~Connes, \emph{{Universal formula for noncommutative
  geometry actions: Unification of gravity and the standard model}},
  \href{https://doi.org/10.1103/PhysRevLett.77.4868}{\emph{Phys. Rev. Lett.}
  {\bfseries 77} (1996) 4868--4871},
  [\href{https://arxiv.org/abs/hep-th/9606056}{{\ttfamily hep-th/9606056}}].

\bibitem{Chamseddine:1996zu}
A.~H. Chamseddine and A.~Connes, \emph{{The Spectral action principle}},
  \href{https://doi.org/10.1007/s002200050126}{\emph{Commun. Math. Phys.}
  {\bfseries 186} (1997) 731--750},
  [\href{https://arxiv.org/abs/hep-th/9606001}{{\ttfamily hep-th/9606001}}].

\bibitem{vanSuijlekom:2024jvw}
W.~D. van Suijlekom, \emph{{Noncommutative Geometry and Particle Physics}}.
\newblock Springer, 12, 2024,
  \href{https://doi.org/10.1007/978-3-031-59120-4}{10.1007/978-3-031-59120-4}.

\bibitem{Chamseddine:2013rta}
A.~H. Chamseddine, A.~Connes and W.~D. van Suijlekom, \emph{{Beyond the
  Spectral Standard Model: Emergence of Pati-Salam Unification}},
  \href{https://doi.org/10.1007/JHEP11(2013)132}{\emph{JHEP} {\bfseries 11}
  (2013) 132}, [\href{https://arxiv.org/abs/1304.8050}{{\ttfamily 1304.8050}}].

\bibitem{Chamseddine:2015ata}
A.~H. Chamseddine, A.~Connes and W.~D. van Suijlekom, \emph{{Grand Unification
  in the Spectral Pati-Salam Model}},
  \href{https://doi.org/10.1007/JHEP11(2015)011}{\emph{JHEP} {\bfseries 11}
  (2015) 011}, [\href{https://arxiv.org/abs/1507.08161}{{\ttfamily
  1507.08161}}].

\bibitem{Aydemir:2018cbb}
U.~Aydemir, D.~Minic, C.~Sun and T.~Takeuchi, \emph{{$B$-decay anomalies and
  scalar leptoquarks in unified Pati-Salam models from noncommutative
  geometry}}, \href{https://doi.org/10.1007/JHEP09(2018)117}{\emph{JHEP}
  {\bfseries 09} (2018) 117},
  [\href{https://arxiv.org/abs/1804.05844}{{\ttfamily 1804.05844}}].

\bibitem{Pati:1974yy}
J.~C. Pati and A.~Salam, \emph{{Lepton Number as the Fourth Color}},
  \href{https://doi.org/10.1103/PhysRevD.10.275,
  10.1103/PhysRevD.11.703.2}{\emph{Phys. Rev.} {\bfseries D10} (1974)
  275--289}.

\bibitem{Mohapatra:1974gc}
R.~N. Mohapatra and J.~C. Pati, \emph{{A Natural Left-Right Symmetry}},
  \href{https://doi.org/10.1103/PhysRevD.11.2558}{\emph{Phys. Rev.} {\bfseries
  D11} (1975) 2558}.

\bibitem{Mohapatra:1974hk}
R.~N. Mohapatra and J.~C. Pati, \emph{{Left-Right Gauge Symmetry and an
  Isoconjugate Model of CP Violation}},
  \href{https://doi.org/10.1103/PhysRevD.11.566}{\emph{Phys. Rev.} {\bfseries
  D11} (1975) 566--571}.

\bibitem{Chang:1984qr}
D.~Chang, R.~N. Mohapatra, J.~Gipson, R.~E. Marshak and M.~K. Parida,
  \emph{{Experimental Tests of New SO(10) Grand Unification}},
  \href{https://doi.org/10.1103/PhysRevD.31.1718}{\emph{Phys. Rev. D}
  {\bfseries 31} (1985) 1718}.

\bibitem{Bertolini:2009qj}
S.~Bertolini, L.~Di~Luzio and M.~Malinsky, \emph{{Intermediate mass scales in
  the non-supersymmetric SO(10) grand unification: A Reappraisal}},
  \href{https://doi.org/10.1103/PhysRevD.80.015013}{\emph{Phys. Rev. D}
  {\bfseries 80} (2009) 015013},
  [\href{https://arxiv.org/abs/0903.4049}{{\ttfamily 0903.4049}}].

\bibitem{Aydemir:2016qqj}
U.~Aydemir and T.~Mandal, \emph{{LHC probes of TeV-scale scalars in
  $\mathrm{SO}(10)$ grand unification}},
  \href{https://doi.org/10.1155/2017/7498795}{\emph{Adv. High Energy Phys.}
  {\bfseries 2017} (2017) 7498795},
  [\href{https://arxiv.org/abs/1601.06761}{{\ttfamily 1601.06761}}].

\bibitem{Aydemir:2019ynb}
U.~Aydemir, T.~Mandal and S.~Mitra, \emph{{Addressing the ${\mathbf
  R_{D^{(*)}}}$ anomalies with an ${\mathbf S_1}$ leptoquark from
  $\mathbf{SO(10)}$ grand unification}},
  \href{https://doi.org/10.1103/PhysRevD.101.015011}{\emph{Phys. Rev. D}
  {\bfseries 101} (2020) 015011},
  [\href{https://arxiv.org/abs/1902.08108}{{\ttfamily 1902.08108}}].

\bibitem{Aydemir:2022lrq}
U.~Aydemir, T.~Mandal, S.~Mitra and S.~Munir, \emph{{An economical model for
  $B$-flavour and $a_\mu$ anomalies from SO(10) grand unification}},
  \href{https://arxiv.org/abs/2209.04705}{{\ttfamily 2209.04705}}.

\bibitem{Chamseddine:2013kza}
A.~H. Chamseddine, A.~Connes and W.~D. van Suijlekom, \emph{{Inner Fluctuations
  in Noncommutative Geometry without the first order condition}},
  \href{https://doi.org/10.1016/j.geomphys.2013.06.006}{\emph{J. Geom. Phys.}
  {\bfseries 73} (2013) 222--234},
  [\href{https://arxiv.org/abs/1304.7583}{{\ttfamily 1304.7583}}].

\bibitem{vanNuland:2021irt}
T.~D.~H. van Nuland and W.~D. van Suijlekom, \emph{{One-loop corrections to the
  spectral action}}, \href{https://doi.org/10.1007/JHEP05(2022)078}{\emph{JHEP}
  {\bfseries 05} (2022) 078},
  [\href{https://arxiv.org/abs/2107.08485}{{\ttfamily 2107.08485}}].

\bibitem{Chamseddine:2012sw}
A.~H. Chamseddine and A.~Connes, \emph{{Resilience of the Spectral Standard
  Model}}, \href{https://doi.org/10.1007/JHEP09(2012)104}{\emph{JHEP}
  {\bfseries 09} (2012) 104},
  [\href{https://arxiv.org/abs/1208.1030}{{\ttfamily 1208.1030}}].

\bibitem{Kurkov:2017wmx}
M.~A. Kurkov and F.~Lizzi, \emph{{Clifford Structures in Noncommutative
  Geometry and the Extended Scalar Sector}},
  \href{https://doi.org/10.1103/PhysRevD.97.085024}{\emph{Phys. Rev.}
  {\bfseries D97} (2018) 085024},
  [\href{https://arxiv.org/abs/1801.00260}{{\ttfamily 1801.00260}}].

\bibitem{Aydemir:2019txw}
U.~Aydemir, \emph{{Clifford-based spectral action and renormalization group
  analysis of the gauge couplings}},
  \href{https://doi.org/10.1140/epjc/s10052-019-6846-9}{\emph{Eur. Phys. J. C}
  {\bfseries 79} (2019) 325},
  [\href{https://arxiv.org/abs/1902.08090}{{\ttfamily 1902.08090}}].

\bibitem{Aydemir:2015nfa}
U.~Aydemir, D.~Minic, C.~Sun and T.~Takeuchi, \emph{{Pati-Salam unification
  from noncommutative geometry and the TeV-scale $W_R$ boson}},
  \href{https://doi.org/10.1142/S0217751X15502231}{\emph{Int. J. Mod. Phys.}
  {\bfseries A31} (2016) 1550223},
  [\href{https://arxiv.org/abs/1509.01606}{{\ttfamily 1509.01606}}].

\bibitem{Aydemir:2016xtj}
U.~Aydemir, D.~Minic, C.~Sun and T.~Takeuchi, \emph{{The 750 GeV diphoton
  excess in unified $SU(2)_L\times SU(2)_R\times SU(4)$ models from
  noncommutative geometry}},
  \href{https://doi.org/10.1142/S0217732316501017}{\emph{Mod. Phys. Lett. A}
  {\bfseries 31} (2016) 1650101},
  [\href{https://arxiv.org/abs/1603.01756}{{\ttfamily 1603.01756}}].

\bibitem{Chamseddine:2019fjq}
A.~H. Chamseddine and W.~D. Van~Suijlekom, \emph{{A survey of spectral models
  of gravity coupled to matter}}.
\newblock \href{https://arxiv.org/abs/1904.12392}{{\ttfamily 1904.12392}}.

\bibitem{BaBar:2012obs}
{ BaBar} collaboration, J.~P. Lees et~al., \emph{{Evidence for an excess of
  $\bar{B} \to D^{(*)} \tau^-\bar{\nu}_\tau$ decays}},
  \href{https://doi.org/10.1103/PhysRevLett.109.101802}{\emph{Phys. Rev. Lett.}
  {\bfseries 109} (2012) 101802},
  [\href{https://arxiv.org/abs/1205.5442}{{\ttfamily 1205.5442}}].

\bibitem{LHCb:2015gmp}
{ LHCb} collaboration, R.~Aaij et~al., \emph{{Measurement of the ratio of
  branching fractions $\mathcal{B}(\bar{B}^0 \to
  D^{*+}\tau^{-}\bar{\nu}_{\tau})/\mathcal{B}(\bar{B}^0 \to
  D^{*+}\mu^{-}\bar{\nu}_{\mu})$}},
  \href{https://doi.org/10.1103/PhysRevLett.115.111803}{\emph{Phys. Rev. Lett.}
  {\bfseries 115} (2015) 111803},
  [\href{https://arxiv.org/abs/1506.08614}{{\ttfamily 1506.08614}}].

\bibitem{Belle:2017ilt}
{ Belle} collaboration, S.~Hirose et~al., \emph{{Measurement of the $\tau$
  lepton polarization and $R(D^*)$ in the decay $\bar{B} \rightarrow D^* \tau^-
  \bar{\nu}_\tau$ with one-prong hadronic $\tau$ decays at Belle}},
  \href{https://doi.org/10.1103/PhysRevD.97.012004}{\emph{Phys. Rev. D}
  {\bfseries 97} (2018) 012004},
  [\href{https://arxiv.org/abs/1709.00129}{{\ttfamily 1709.00129}}].

\bibitem{Crivellin:2025txc}
A.~Crivellin and B.~Mellado, \emph{{Anomalies in Particle Physcis}},
  \href{https://doi.org/10.22323/1.469.0007}{\emph{PoS} {\bfseries DIS2024}
  (2025) 007}.

\bibitem{HeavyFlavorAveragingGroupHFLAV:2024ctg}
{ Heavy Flavor Averaging Group (HFLAV)} collaboration, S.~Banerjee et~al.,
  \emph{{Averages of $b$-hadron, $c$-hadron, and $\tau$-lepton properties as of
  2023}},  \href{https://arxiv.org/abs/2411.18639}{{\ttfamily 2411.18639}}.

\bibitem{Bauer:2015knc}
M.~Bauer and M.~Neubert, \emph{{Minimal Leptoquark Explanation for the
  $R_{D^{(*)}}$, $R_K$, and $(g-2)_\mu$ Anomalies}},
  \href{https://doi.org/10.1103/PhysRevLett.116.141802}{\emph{Phys. Rev. Lett.}
  {\bfseries 116} (2016) 141802},
  [\href{https://arxiv.org/abs/1511.01900}{{\ttfamily 1511.01900}}].

\bibitem{Angelescu:2018tyl}
A.~Angelescu, D.~Be?irevi?, D.~A. Faroughy and O.~Sumensari, \emph{{Closing the
  window on single leptoquark solutions to the $B$-physics anomalies}},
  \href{https://doi.org/10.1007/JHEP10(2018)183}{\emph{JHEP} {\bfseries 10}
  (2018) 183}, [\href{https://arxiv.org/abs/1808.08179}{{\ttfamily
  1808.08179}}].

\bibitem{Aydemir:2023sty}
U.~Aydemir, \emph{{$B$-flavour and $a_\mu$ anomalies with $S_1$ leptoquark in
  SO(10) Grand Unification}},  in \emph{{Beyond Standard Model: From Theory to
  Experiment}}, 2023, \href{https://doi.org/10.31526/BSM-2023.24}{DOI}.

\bibitem{Buchmuller:1986zs}
W.~Buchmuller, R.~Ruckl and D.~Wyler, \emph{{Leptoquarks in Lepton - Quark
  Collisions}}, \href{https://doi.org/10.1016/0370-2693(87)90637-X}{\emph{Phys.
  Lett. B} {\bfseries 191} (1987) 442--448}.

\bibitem{Dorsner:2016wpm}
I.~Dor{\v{s}}ner, S.~Fajfer, A.~Greljo, J.~F. Kamenik and N.~Ko{\v{s}}nik,
  \emph{{Physics of leptoquarks in precision experiments and at particle
  colliders}}, \href{https://doi.org/10.1016/j.physrep.2016.06.001}{\emph{Phys.
  Rept.} {\bfseries 641} (2016) 1--68},
  [\href{https://arxiv.org/abs/1603.04993}{{\ttfamily 1603.04993}}].

\bibitem{Nath:2006ut}
P.~Nath and P.~Fileviez~Perez, \emph{{Proton stability in grand unified
  theories, in strings and in branes}},
  \href{https://doi.org/10.1016/j.physrep.2007.02.010}{\emph{Phys. Rept.}
  {\bfseries 441} (2007) 191--317},
  [\href{https://arxiv.org/abs/hep-ph/0601023}{{\ttfamily hep-ph/0601023}}].

\bibitem{Mohapatra:1980qe}
R.~N. Mohapatra and R.~E. Marshak, \emph{{Local B-L Symmetry of Electroweak
  Interactions, Majorana Neutrinos and Neutron Oscillations}},
  \href{https://doi.org/10.1103/PhysRevLett.44.1316}{\emph{Phys. Rev. Lett.}
  {\bfseries 44} (1980) 1316--1319}.

\bibitem{Bhaskar:2022vgk}
A.~Bhaskar, A.~A. Madathil, T.~Mandal and S.~Mitra, \emph{{Combined explanation
  of W-mass, muon g-2, RK(*) and RD(*) anomalies in a singlet-triplet scalar
  leptoquark model}},
  \href{https://doi.org/10.1103/PhysRevD.106.115009}{\emph{Phys. Rev. D}
  {\bfseries 106} (2022) 115009},
  [\href{https://arxiv.org/abs/2204.09031}{{\ttfamily 2204.09031}}].


\end{thebibliography}
\end{document}